\documentclass[twocolumn,twoside,slac_two]{revtex4}
\usepackage{graphicx}
\usepackage{fancyhdr}
\usepackage{units}
\usepackage{url}
\usepackage{subfigure}
\newcommand*\aap{A\&A}

\newcommand*\aapr{A\&A~Rev.}

\newcommand*\apj{ApJ}
\newcommand*\apjl{ApJ}

\newcommand*\apjs{ApJS}

\newcommand*\apss{Ap\&SS}

\newcommand*\mnras{MNRAS}

\newcommand{\lsi}{LS~I~+61$^{\circ}$303}

\begin{document}

\title{The apparent discontinuity in the periodicity of the GeV emission from \lsi{}}


\author{F.~Jaron, M.~Massi}
\affiliation{Max-Planck-Institute for Radio Astronomie, Auf dem H\"ugel
  69, D-53121 Bonn, Germany}

\begin{abstract}
	The $\gamma$-ray binary\ \lsi{} shows a discontinuity of the periodicity
in its GeV emission. 
%We used a wavelet analysis to explore the temporal development of periodic
%signals. The wavelet analysis was first applied to the whole data set
%of available \textit{Fermi}-LAT data and then to the two subsets of orbital
%phase intervals $\Phi = 0.0-0.5$ and $\Phi = 0.5-1.0$. We also performed a
%Lomb-Scargle timing analysis. We investigated the similarities between GeV
%$\gamma$-ray emission and radio emission by comparing the folded curves of the
%\textit{Fermi}-LAT data and the Green Bank Interferometer radio data.
In this paper, we show that during the epochs when the timing analysis fails to determine the orbital
periodicity, the periodicity is in fact present in the two orbital phase intervals
$\Phi = 0.0-0.5$ and $\Phi = 0.5-1.0$. That is, there are two periodic
signals, one towards periastron (i.e., $\Phi = 0.0-0.5$) and another one towards
apastron ($\Phi = 0.5-1.0$). The apastron peak shows the same
orbital shift as the radio outburst and, in addition, reveals the same two
periods $P_1$ and $P_2$ that are present in the radio data.
The $\gamma$-ray emission of the apastron peak normally just broadens
the emission of the peak around periastron. Only when it appears at $\Phi =
0.8-1.0$ because of the orbital shift, it is detached enough from the first
peak to become recognizable as a second orbital peak, which is the reason why
the timing analysis fails. Two $\gamma$-ray peaks along the orbit are predicted
by the two-peak accretion model for an eccentric orbit that was proposed
by several authors for \lsi{}.
\end{abstract}

\maketitle

\thispagestyle{fancy}

\section{Introduction}

\begin{figure}
	\includegraphics[height=\columnwidth, angle=-90]{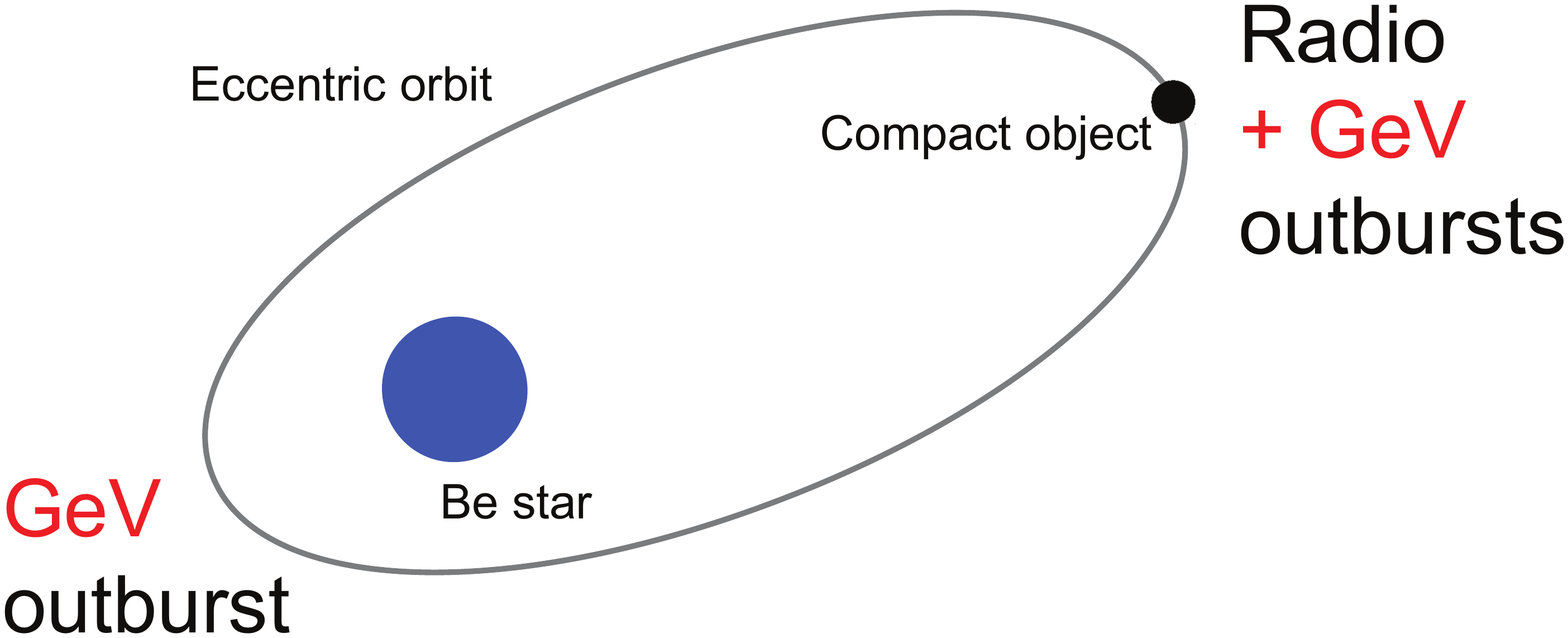}
	\caption{Sketch of \lsi{}. The periodic GeV outburst at periastron was first reported by \cite{Abdo2009}. In Sect.~3 a newly discovered periodic apastron GeV peak is discussed \cite{Jaron2014}. The radio outburst occurs only at apastron, whereas at periastron, only a low level of emission is present \cite[see Fig.~1-Right in][]{Massi2015}.}
	\label{fig1}
\end{figure}

The stellar system \lsi{} is a member of the
small class of $\gamma$-ray binaries, which are defined as binary stars with
a peak in the spectral energy distribution above 1~MeV \cite{Dubus2013}.
A sketch of the system is shown in Fig.~\ref{fig1}. \lsi{} consists of a Be star and a compact object in an
eccentric orbit, $e = 0.72 \pm 0.15$ \cite{Casares2005}, with orbital
period $P_1 = \unit[26.4960 \pm 0.0028]{d}$ \cite{Gregory2002}. The Be star is
rapidly rotating and losing mass in form of an equatorial disk
\cite{Casares2005}. The nature of the compact object could not yet be
established, because the masses are poorly constrained due to the large
uncertainty in the inclination angle \cite{Casares2005}. The orbital phase of
the binary system is defined as
\begin{equation}
  \Phi = \frac{t - t_0}{P_1} - {\rm int}\left(\frac{t - t_0}{P_1}\right),
\end{equation}
where $t_0 = {\rm MJD}~43366.275$ \cite{Gregory2002}. Periastron occurs at
orbital phase $\Phi = 0.23$ \cite{Casares2005}.

Radio outbursts are observed at
orbital phases $\Phi = 0.5-0.9$, i.e., around apastron. Their peak flux
densities are modulated in amplitude and orbital phase occurrence by a
long-term period $P_{\rm long} = \unit[1667 \pm 8]{d}$ \cite{Gregory2002}. The
long-term phase\ $\Theta$ is defined analogous to the orbital phase $\Phi$ by
replacing $P_1$ by $P_{\rm long}$.

%Consecutive VLBI images of the source covering one orbital cycle show an
%elongated structure which is sometimes one-sided and sometimes two-sided,
%changing position angle along the orbit with a precessional period of
%27--28~days close to the orbital period \cite{Massi2012}. A fast precessing
%relativistic jet would give rise to variable Doppler boosting, and indeed
%Lomb-Scargle timing analysis of GBI radio data at 2~GHz and 8~GHz results in
%two highly significant periods, $P_1 = \unit[26.49 \pm 0.07]{d}$ and $P_2
%= \unit[26.92 \pm 0.07]{d}$ (see Fig.~\ref{fig:radioSC}). The first period\
%$P_1$ coincides with the previously determined orbital period
%\cite{Gregory2002}, while the second, $P_2$, agrees well with the precessional
%period derived from VLBI images \cite{Massi2012}. In addition, these two close
%periodicities give rise to a beating with period $P_{\rm beat} = \left(P_1^{-1}
%- P_2^{-1}\right)^{-1} = \unit[1667 \pm 393]{d}$, which explains both the
%long-term modulation in amplitude and orbital phase occurrence of the radio
%outbursts \cite{Massi2013}.

The source \lsi{} is highly variable and periodic all over the electromagnetic
spectrum from radio to very high energy $\gamma$-rays \cite{Gregory2002,
Casares2005, Abdo2009, Albert2009}. 
The GeV $\gamma$-ray light curve, as obtained using \textit{Fermi} LAT data, has
so far been reported to peak at orbital phases around periastron \cite{Abdo2009,
Hadasch2012} (see Fig.~\ref{fig1}). Timing analysis shows that
the orbital period is present in the \textit{Fermi} LAT light
curve from this source, however not with equal power all of the time
\cite{Hadasch2012, Ackermann2013}.
There are times ($\Theta$-phases) when the period is outstanding and there are times when the period is completely absent from the power spectrum, as shown well in Fig.~4 of \cite{Ackermann2013}. 
Moreover, Fig.~3 of \cite{Ackermann2013} shows that GeV data also show the long-term periodical
variation affecting the radio data, but only at a
specific orbital phase interval, $\Phi = 0.5-1.0$, that is around
apoastron.

We are aimed here to investigate the discontinuity
in the periodicity of the GeV 
$\gamma$-ray emission at periastron,
the possible relationship of its disappearance with the variation of
the emission around apastron, and finally the
possible relationship between GeV and radio emission.

\section{Data analysis}

For the present analysis \cite{Jaron2014} we use \textit{Fermi} LAT data from \lsi{} spanning
the time range August 5, 2008 (MJD~54683) until June~30, 2014 with an energy
range of 100~MeV to 300~GeV. For the
computation of the light curves we used the script like\_lc.pl written by
Robin~Corbet.
\footnote{\url{http://fermi.gsfc.nasa.gov/ssc/data/analysis/user/}}
Only source-event-class photons were selected for the analysis. Photons with
a zenith angle greater than 100$^{\circ}$ were excluded to reduce contamination
from the Earth's limb. For the diffuse emission we used the model
gll\_iem\_v05\_rev1.fit and the template iso\_source\_v05\_rev1.txt. We used the
instrument response function (IRF) P7REP/background\_rev1, and the model file
was generated from the 2FGL catalogue \cite{Nolan2012}, all sources within
$10^{\circ}$ of \lsi{} were included in the model. \lsi{} was fitted with a
log-parabola spectral shape and with all parameters left free for the fit,
performing an unbinned maximum likelihood analysis. The other sources were fixed
to their catalogue values. We produced light curves with a time bin size of one
day and of five days.

We investigated \cite[for details see][]{Jaron2014} the temporal evolution of the orbital periodicity by means of a
wavelet analysis \cite{Torrence1998} and Lomb-Scargle timing analysis \cite{Lomb1976, Scargle1982}. 

\section{Results: A periodic signal around apastron}

\subsection{Wavelet analysis}

\begin{figure*}[h]
  \subfigure[]{
    \hbox{\hspace{-.66em}
      \includegraphics{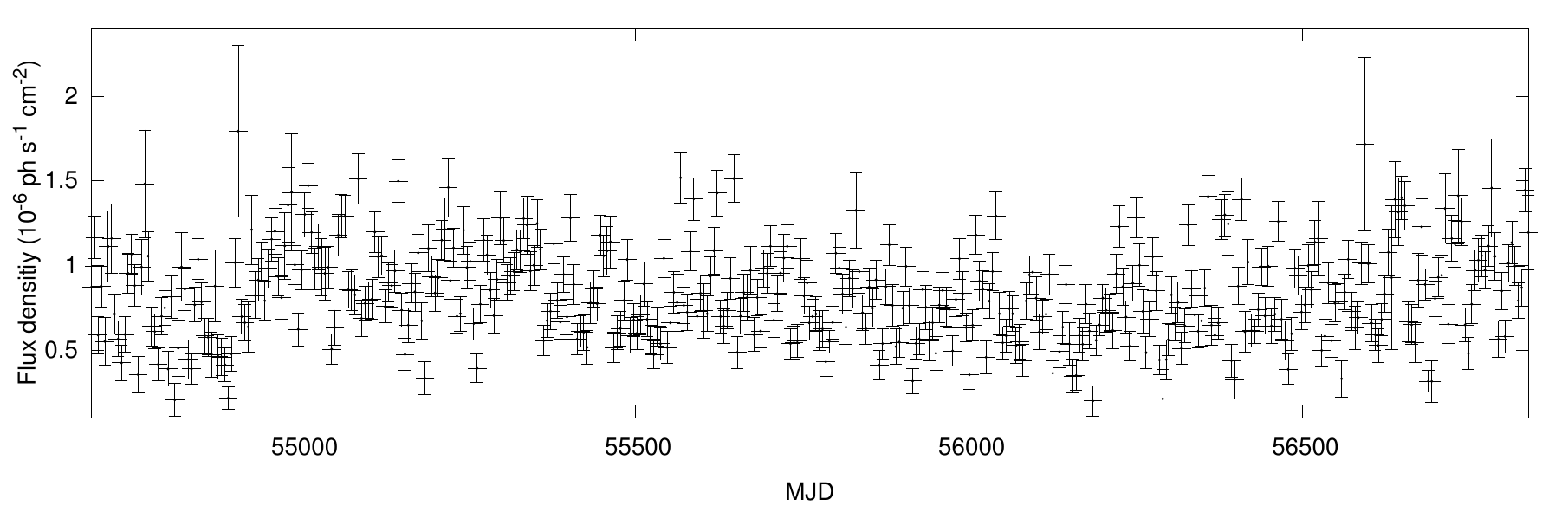}
    }
  }
  \subfigure[]{
    \includegraphics{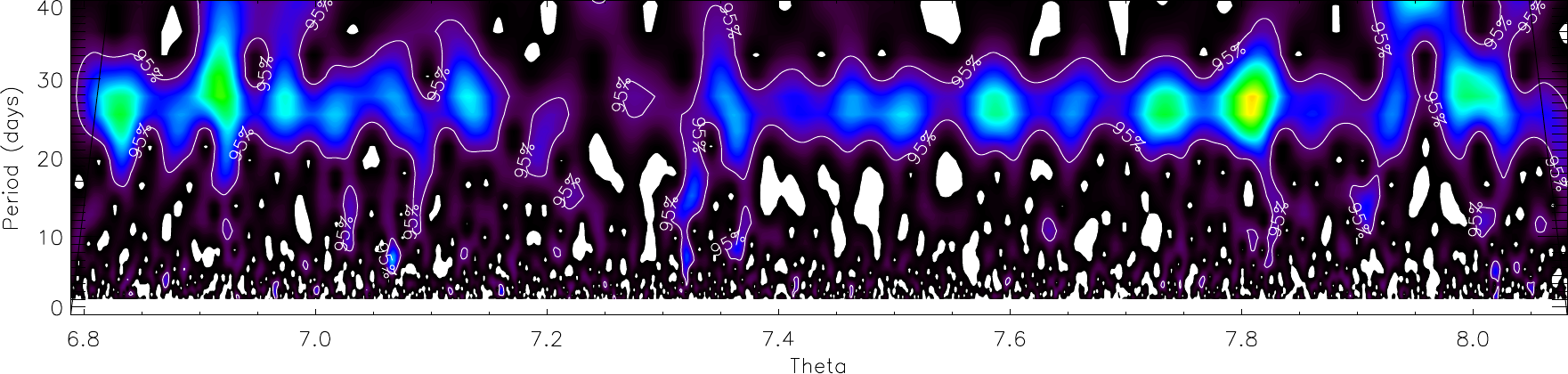}
  }
  \subfigure[]{
    \includegraphics{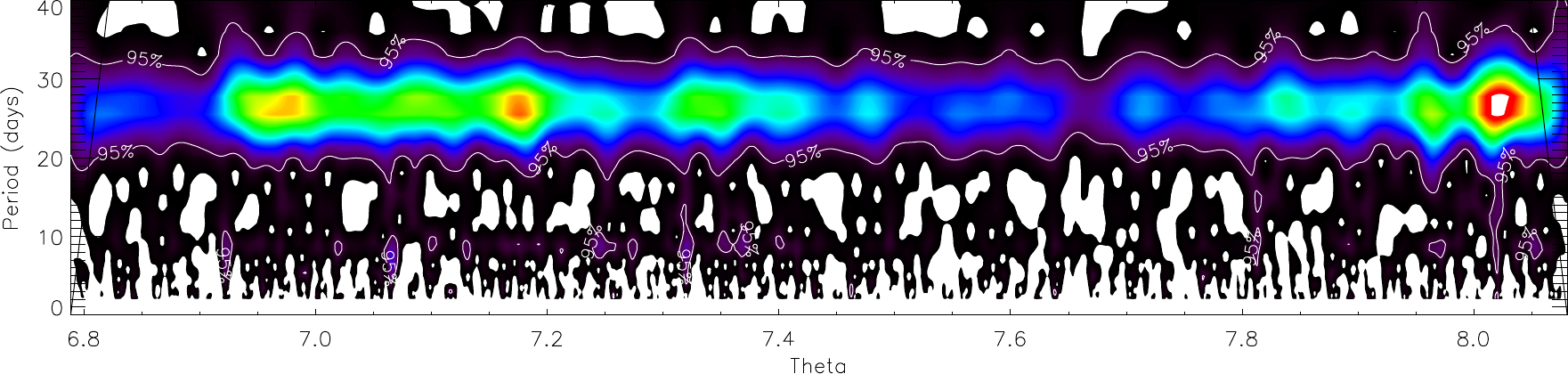}
  }
  \subfigure[]{
    \includegraphics{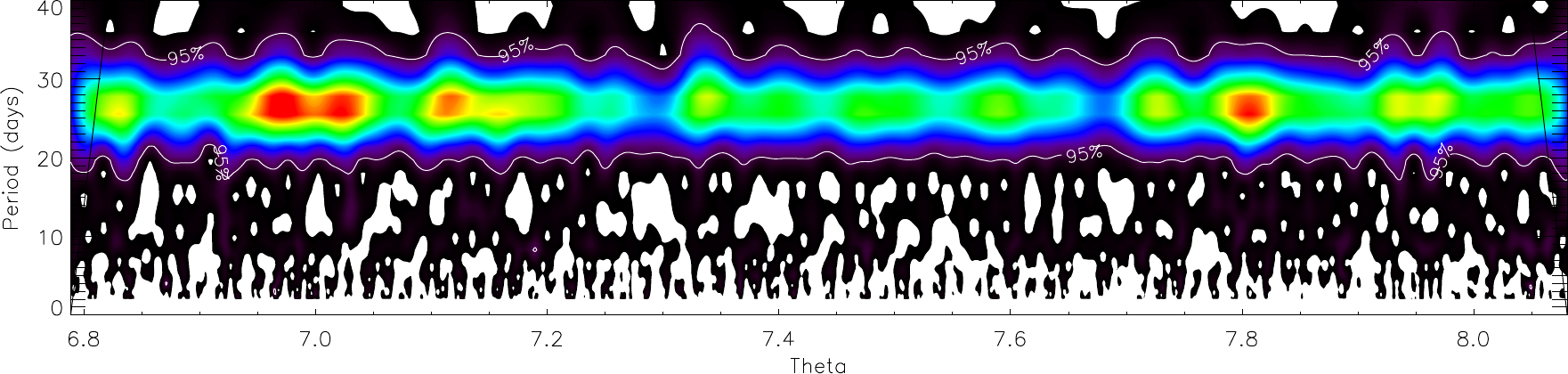}
  }
  \subfigure{
    \includegraphics{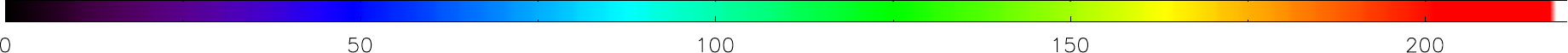}
  }
  \caption{
    Wavelet analysis of \textit{Fermi}-LAT data. The strength of
    periodicity is colour coded as indicated in the bottom bar.
    (a) \textit{Fermi}-LAT data with a time bin of 1 d.
    (b) Wavelet analysis for the whole orbital
	interval $0.0-1.0$ (b--d use a time bin of one day).
    (c) Wavelet analysis for the orbital
	interval\ $\Phi = 0.5 -1.0$, i.e., around apoastron.
    (d) Wavelet for the orbital interval\ $\Phi
	= 0.0 -0.5$, i.e., around periastron.
  }
  \label{fig:FermiLATWL}
\end{figure*}

Our results are shown Fig.~2.
The first plot of Fig.~2 presents the examined data set.
The wavelet analysis was applied to the $\gamma$-ray data vs time, however, for a
straightforward comparison with radio data, we express in the other plots of Fig.~2 the $x$-axis as the
long-term phase\ $\Theta$. 
%This allows a comparison with non-simultaneous radio
%data because the radio data are periodic in $\Theta$. 
%We will therefore compare
%$\gamma$-ray data to radio data having the same fractional part of $\Theta$.
The second plot of Fig.~\ref{fig:FermiLATWL} shows the wavelet plot for the whole data set, i.e., the whole orbital period $\Phi = 0.0-1.0$.
%of
%the entire \textit{Fermi} LAT data from \lsi{}. 
The absence of
the orbital period around $\Theta \approx 7.2$ is consistent with the previous
finding shown in Fig.~4 of \cite{Ackermann2013}. When wavelet analysis is
performed only on data from the orbital phase intervals $\Phi = 0.0 - 0.5$
(middle) and $\Phi = 0.5 - 1.0$ (bottom), it is revealed that \emph{there is always a
periodic signal at $\Phi = 0.0 - 0.5$ (periastron). Moreover, there is a
periodic signal at $\Phi = 0.5 - 1.0$ (apastron)}. The latter becomes
particularly strong during the time when the orbital period is absent from the
power spectra of $\Phi = 0.0-1.0$ \cite{Jaron2014}.

\subsection{Lomb-Scargle timing analysis}

\begin{figure*}
  \includegraphics[width=\textwidth]{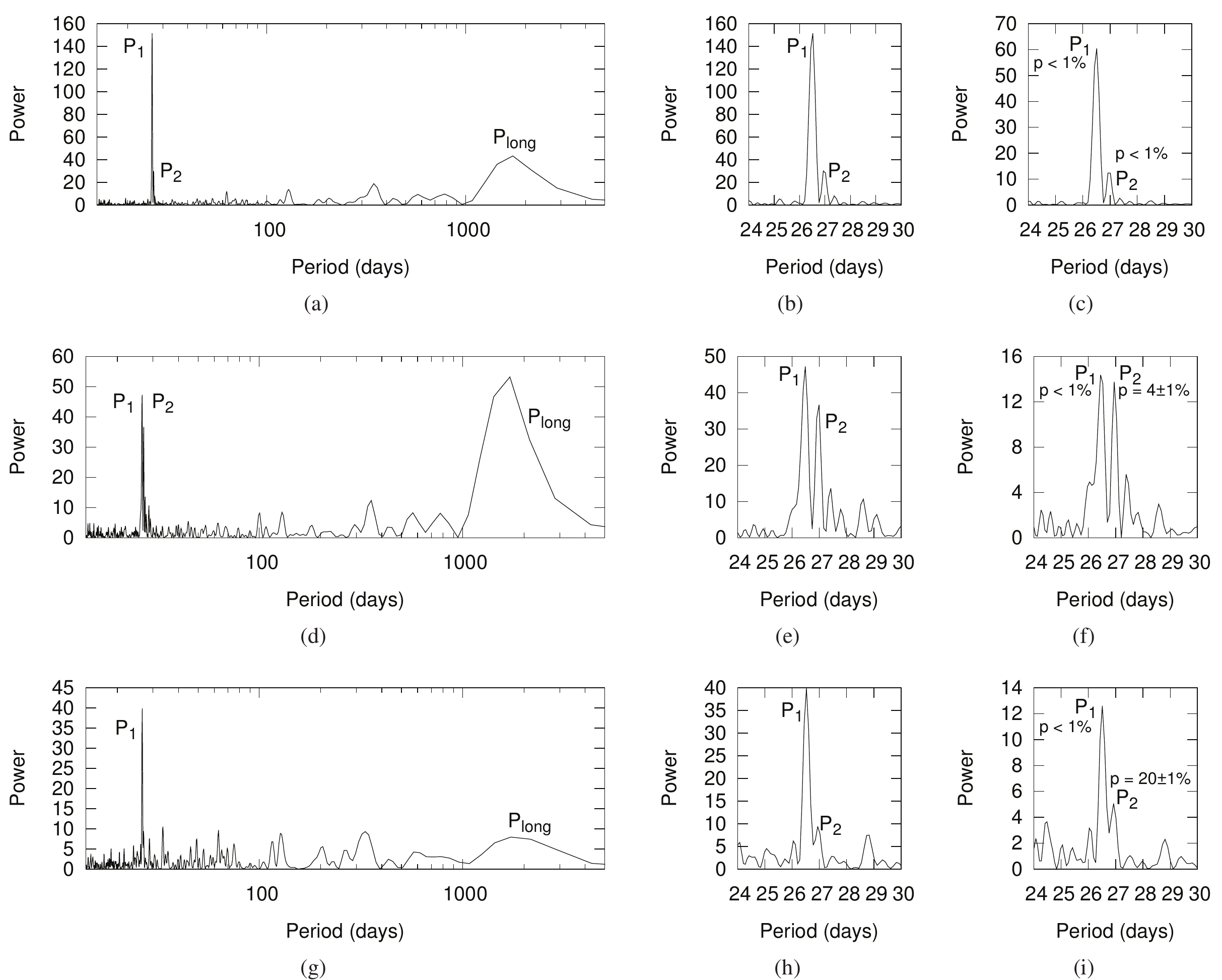}
  \caption{
    Lomb-Scargle periodogram of the \textit{Fermi} LAT data (with a time bin of
    one day). Figure 3 in \cite{Jaron2014}. 
    (a) Data in the orbital phase $\Phi = 0.0 - 1.0$. 
    (b) Zoom of Fig.~\ref{fig:Jaron2014_2}\,a. 
    (c) Same as \ref{fig:Jaron2014_2}\,b for data with a time bin of 5~d. 
    (d) Data in the orbital phase $\Phi = 0.5 - 1.0$. The periods $P_2$ and
	$P_{\rm long}$ here present are typical periodicities in radio data
	\cite{Massi2013}. 
    (e) Zoom of Fig.~\ref{fig:Jaron2014_2}\,d. 
    (f) Same as \ref{fig:Jaron2014_2}\,e for data with a time bin of 5~d. 
    (g) Data in the orbital phase $\Phi = 0.0 - 0.5$. 
    (h) Zoom of Fig.~\ref{fig:Jaron2014_2}\,g.
    (i) Same as \ref{fig:Jaron2014_2}\,h for data with a time bin of 5~d.
  }
  \label{fig:Jaron2014_2}
\end{figure*}

Figure~\ref{fig:Jaron2014_2} shows Lomb-Scargle periodograms of the $\gamma$-ray
flux from \lsi{}. The data have been selected from orbit phase intervals like in
the previous section. In the periodogram for the entire
orbit (Fig.~\ref{fig:Jaron2014_2}\,a) the strongest feature is a peak which
agrees well with the orbital period $P_1$ found by \cite{Gregory1999}. 
%The zoom
%in Fig.~\ref{fig:Jaron2014_2}\,b reveals that there is a small yet significant
%second peak at a period in agreement with the previously found precessional
%period \cite{Massi2013} (see also Fig.~\ref{fig:radioSC}). This second peak is
%also present in the data integrated over five days, shown
%in Fig.~\ref{fig:Jaron2014_2}\,c. 
Figures~\ref{fig:Jaron2014_2}\,d, e, and f, refer to only data from $\Phi = 0.5-1.0$. In this orbital phase interval the peak at $P_{\rm long}$ is a very
strong feature, in agreement with the findings of \cite{Ackermann2013}.
Moreover, the zoom of Fig.~\ref{fig:Jaron2014_2}\,d, i.e., Fig.~\ref{fig:Jaron2014_2}\,e, shows a second peak, $P_2= \unit[26.99 \pm
0.08]{d}$.
This second peak becomes
stronger and is almost as strong as the peak at $P_1 = \unit[26.48 \pm 0.08]{d}$ in the 5 day integrated
data in Fig.~\ref{fig:Jaron2014_2}\,f. 
The periods $P_1$,  $P_2$ (see Fig.~4), and $P_{\rm long}$ here present are typical
periodicities in radio data as shown in \cite{Massi2013}.

\subsection{Folded \textit{Fermi} LAT data: The apastron GeV peak and its
orbital shift}

\begin{figure*}
  \includegraphics[height=\textwidth, angle=-90]{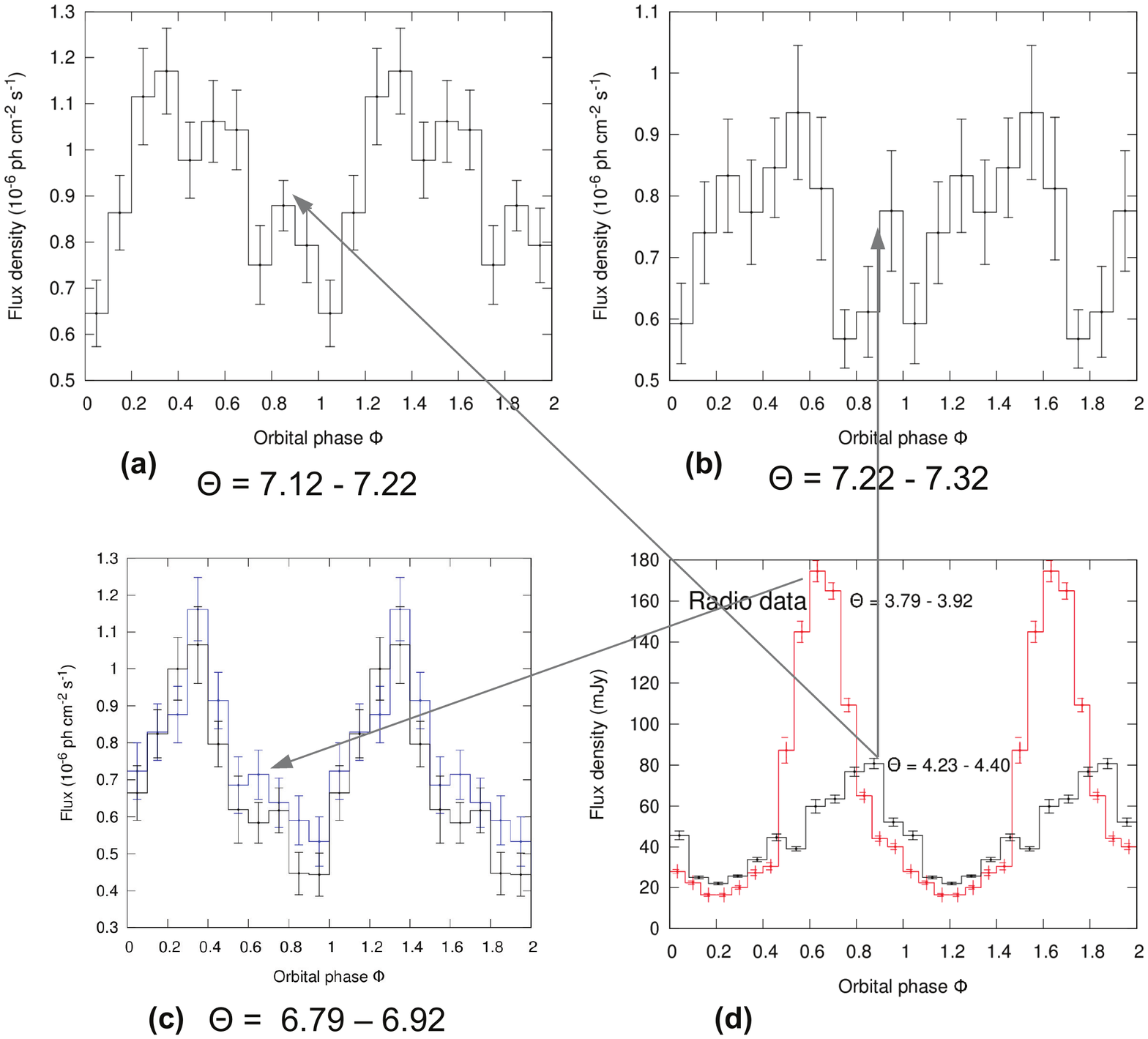}
  \caption{
	  (a)-(c) Folded \textit{Fermi} LAT $\gamma$-ray data (100~MeV -- 300~GeV). The blue curve in (c) is that of \cite{Hadasch2012}. 
    (d) Folded GBI 8~GHz radio data.
    The here discovered periodic apastron GeV peak follows the same timing
    characteristic (i.e., $P_1$ and $P_2$ are both present) as the periodic radio peak, which also occurs around
    apastron. During the time when the orbital period disappears from the power
    spectra of the $\gamma$-ray light curve (see Fig.~2\,b, $\Theta \approx 7.2$)
    the apastron GeV peak becomes well visible in the folded light curve, because it is more displaced from the periastron peak (see \cite{Jaron2014} and here Sect.~4).
  }
  \label{fig:Jaron_poster_1}
\end{figure*}

Figure~5\,a and b show \textit{Fermi} LAT data from the time ($\Theta \approx 7.2$) of the disappearance of the orbital period from the power spectra folded with the orbital period. A second peak is evident at orbital phases $\Phi = 0.8-1.0$. Figure~5\,c shows \textit{Fermi} LAT data for another $\Theta$. It is now interesting to compare these plots with radio data. Because of the long-term periodicity we can compare $\gamma$-ray and radio data having the same fractional part of $\Theta$. Figure~5\,d shows GBI radio data at 8~GHz \cite[for details see][]{Jaron2014}. 

\section{Conclusions}

During the intervals where the orbital periodicity is absent from the
power spectra, wavelet and the folded light curves show two periodic signals,
one at periastron and a second at apastron. The presence of
the second periodic outburst disturbs the timing analysis and prevents it
from finding the orbital periodicity. Comparison with the folded radio data
(Fig.~4\,d) suggests that the apastron GeV peak follows the same orbital shift as the
radio outbursts \citep{Jaron2014}. It is well-known the
phenomenon of the orbital shift of the radio outburst in
\lsi{}: The largest outbursts occur at orbital phase
0.6, afterwards, with the long-term periodicity, the orbital
phase of the peak of the outburst changes, as analysed by
\cite{Paredes1990} in terms of orbital phase shift, by
\cite{Gregory1999} in terms of timing residuals, and reproduced
recently by the precessing jet model in \cite{Massi2014}, here shown in Fig.~5.

Our result of two
GeV peaks along the orbit corroborates the two-peak
accretion model for \lsi{}. The hypothesis that
a compact object that accretes material along an eccentric
orbit undergoes two accretion peaks along the orbit
was suggested and developed by several authors for the
system \lsi{} \cite{Taylor1992, Marti1995, Bosch-Ramon2006, Romero2007}.
The first accretion peak is predicted to occur close to the
Be star and to give rise to a major high-energy outburst.
The second accretion peak is predicted to occur much
farther away from the Be star, where the radio outburst
occurs, and a minor high-energy outburst is predicted
there \cite{Bosch-Ramon2006}. The predicted periastron
event corresponds well to the observed GeV peak
towards periastron, the second predicted high-energy
outburst, corresponds well to the here discussed apoastron
peak.

\begin{figure*}
	\includegraphics[width=\columnwidth]{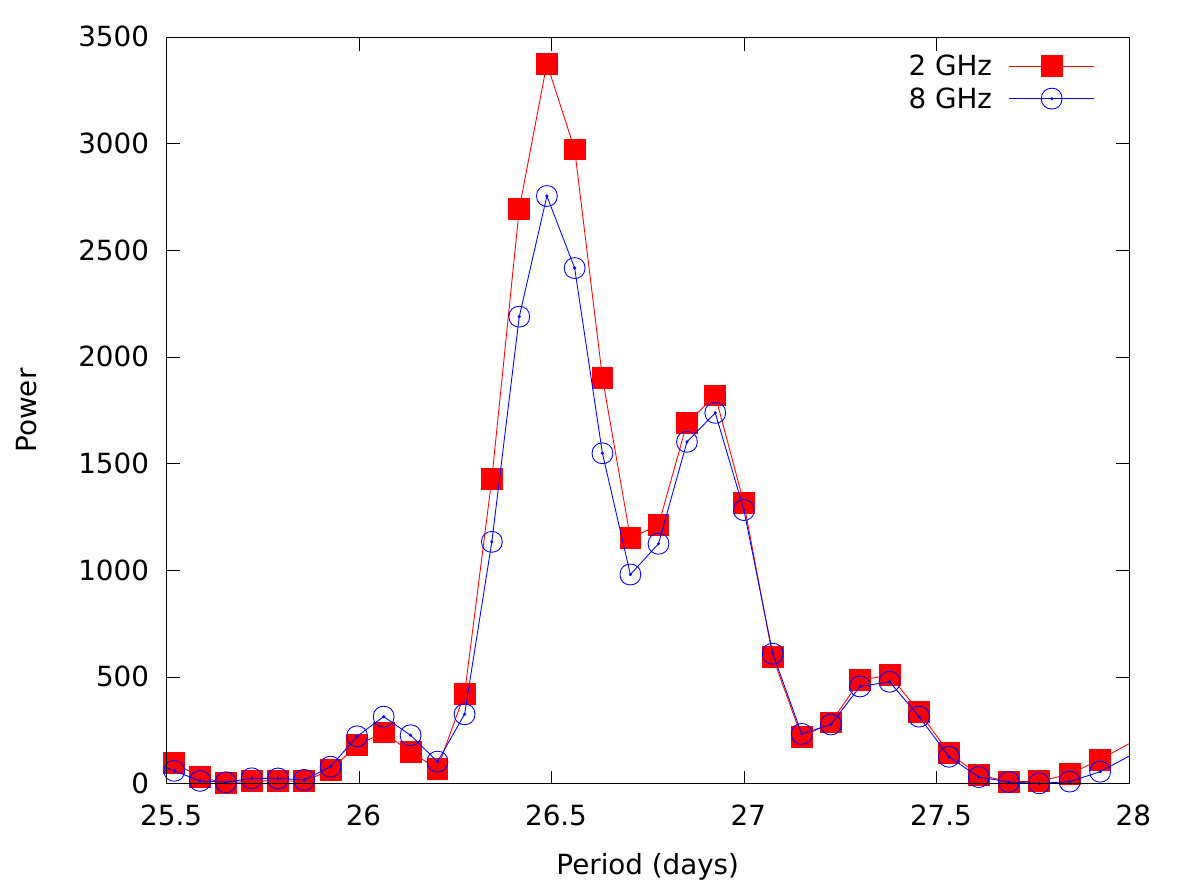}
	\includegraphics[width=\columnwidth]{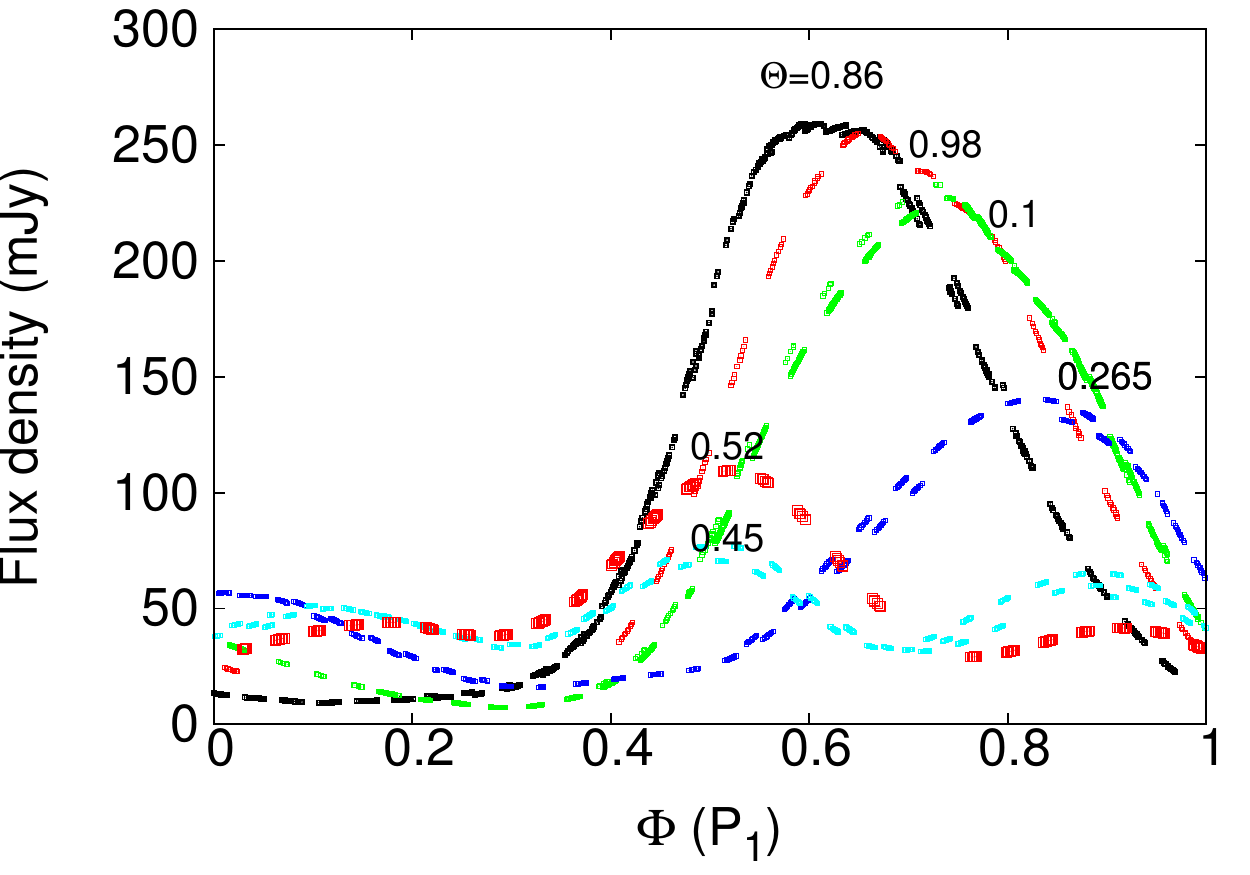}
  \caption{
 Left: Timing analysis of 6.7~years of GBI radio data at 2 and 8~GHz results
    in two periods, $P_1 = \unit[26.49 \pm 0.07]{d}$, $P_2 = \unit[26.92 \pm
    0.07]{d}$. The long-term period $P_{\rm long} = \unit[1667 \pm 8]{d}$ is
    consistent with the period $P_{\rm beat} = 1/(\nu_1-\nu_2) = \unit[1667 \pm
    393]{d}$ resulting from the beating between the two close periodicities
    $P_1$ and $P_2$ \cite{Massi2013}.
Right: Orbital shift of the radio outburst of \lsi{} in the precessing jet model of \cite{Massi2014}. At $\Theta = 0.86$ the outbursts peak at $\Phi \approx 0.6$. At $\Theta = 0.265$ the outbursts peak at $\Phi \approx 0.85$.}
	\label{fig:radio}
\end{figure*}

\bigskip % extra skip inserted
\begin{acknowledgments}
	We thank Bindu~Rani for carefully reading the manuscript and for useful comments.
We thank Robin~Corbet for answering our questions concerning the computation of
light curves. We thank Walter~Alef and Alessandra~Bertarini for their
assistance with computation power.
Wavelet software was provided by C.~Torrence and G.~Compo, and is available at
URL: \url{http://atoc.colorado.edu/research/wavelets/}. The Green Bank
Interferometer was operated by the National Radio Astronomy Observatory for
the U.S. Naval Observatory and the Naval Research laboratory during
the time period of these observations. This work has made use of public
\textit{Fermi} LAT data obtained from the High Energy Astrophysics Science
Archive Research Center (HEASARC), provided by NASA Goddard Space Flight
Center.
\end{acknowledgments}

\bigskip % extra skip inserted

\end{document}